\documentclass[12pt,a4paper]{article}
\usepackage[T2A]{fontenc}
\usepackage[utf8]{inputenc}
\usepackage[english]{babel}
\usepackage{amssymb}
\usepackage{amsmath}
\usepackage{graphicx}
\usepackage[small,sc,center]{titlesec}
\usepackage{indentfirst}
\usepackage{siunitx}
\usepackage[labelsep=period]{caption}
\usepackage{caption}

\newcommand{\nc}{\newcommand}
\nc{\etab}{\eta_\textrm{B}}
\nc{\Teff}{T_\textrm{eff}}
\nc{\tev}{t_\textrm{ev}}
\begin{document}

\begin{center}
	\textbf{Hydrodynamic model of decaying radial oscillations in RU Cam}
	
	\vskip 3mm
	\textbf{Yu. A. Fadeyev\footnote{E--mail: fadeyev@inasan.ru}}
	
	\textit{Institute of Astronomy, Russian Academy of Sciences,
		Pyatnitskaya ul. 48, Moscow, 119017 Russia} \\
	
	Received October 6, 2021; accepted November 3, 2021
\end{center}

\textbf{Abstract} ---
Calculations of stellar evolution up to the early white dwarf stage were carried out for
stars with mass on the main sequence $M_0=0.82M_\odot$, $0.85M_\odot$, $0.9M_\odot$ and
with initial abundances of helium and heavier elements $Y=0.25$ and $Z=10^{-3}$, respectively.
For each value of $M_0$ the AGB and post--AGB evolutionary phases were computed with three
values of the mass loss parameter in the Bl\"ocker formula: $\etab=0.02$, 0.05 and 0.1.
The variable star RU~Cam with pulsation period $\Pi\approx 22$ day is shown to be in the
post--AGB stage and the pulsation amplitude decrease in years 1962--1963 is due to
movement of the star across the HR diagram beyond the pulsation instability region.
Theoretical estimates of the mass and the luminosity of RU~Cam are
$0.524M_\odot\le M\le 0.532M_\odot$ and
$2.20\times 10^3L_\odot\le L\le 2.33\times 10^3L_\odot$, respectively.
Hydrodynamic calculations of nonlinear stellar pulsations show that while the star approaches
the instability boundary a significant reduction ($\approx 90\%$) in the pulsation amplitude
occurs for nearly two years with subsequent slow decay of low--amplitude oscillations.
Solution of the equations of hydrodynamics with time--dependent inner boundary conditions
describing evolutionary changes in the radius and the luminosity at the bottom of the
pulsating envelope allows us to conclude that decay of radial oscillations in RU~Cam
is accompanied by the effect of oscillation hysteresis.
In particular, the stage of large--amplitude limit cycle oscillations extends by
$\approx 12$ years and the subsequent stage of decaying small--amplitude oscilations
spreads beyond the formal boundary of pulsation instability.


\newpage
\section*{introduction}

The variable star RU~Cam (BD$+69^\circ 417$) with pulsation period $\Pi\approx 22$~day
belongs to W~Vir pulsating variables (Samus' et al. 2017).
As in many variable stars of this type the period and the shape of the light curve
of RU~Cam are unstable (Nielsen 1936; Payne--Gaposchkin 1941; Lenouvel and Jehoulet 1953).
Nevertheless, the $O-C$ diagram based on the published photometric observations from 1907
to 1965 allows us to conclude that at present the pulsation period of this star undergoes
secular decrease (Szeidl et al. 1992).
The average light of RU~Cam remains approximately the same therefore the period
decrease means evolutionary movement of the star across the Hertzsprung--Russel (HR)
diagram to higher effective temperatures.

RU~Cam has attracted great attention after the report on abrupt decrease of its
pulsation amplitude (Demers and Fernie 1966).
Before 1963 the amplitude of light changes was $\Delta m_{pg}\approx 1$~mag
(Ceraski 1907; Sanford 1928; Payne--Gaposchkin 1941; Lenouvel and Jehoulet 1953)
with radial velocity amplitude $\Delta V_r \approx 30$~km/s (Sanford 1928).
However, after 1966 the amplitudes reduced to $\Delta m_V < 0.2$~mag
(Demers and Fernie 1966; Broglia 1967; Fernie 1968; Broglia and Guerrero 1972;
Berdnikov and Voziakova 1995) and $\Delta V_r < 10$~km/s (Wallerstein and Crampton 1967).
The last photoelectric light measurements of RU~Cam before the pulsation amplitude drop
were made from January to April of 1962 (Michalowska-Smak and Smak 1965), whereas the
agerage light curve with amplitude nearly 0.2~mag is based on observations caried out
from October 1964 to January 1966 (Demers and Fernie 1966).
Therefore, the amplitude decline was no longer than two years.
The chronology of RU~Cam period and amplitude changes is presented in more detail by
Percy (1921).

The origin of rapid pulsation amplitude decline in RU~Cam is still unclear.
Whitney (1967) supposed that the abrupt cessation of pulsation is due to the effect
of oscillation hysteresis when the star crosses the boundary of pulsation instability.
The oscillation hysteresis appears in the slowly evolving non--linear system so that
the large--amplitude limit--cycle oscillations are sustained when the system occurs
in conditions of stability against the small-amplitude oscillations with subsequent
abrupt cessation of oscillations (Bogoliubov and Mitropolsky 1961).
Unfortunately, a quantitative model of oscillation hysteresis in stellar pulsations
has not been developed yet.
Koll\'ath and Szeidl (1993) supposed that abrupt decrease of the amplitude is due to
cessation of regular large--amplitude oscillations whereas the small--amplitude
light variability observed later is due to irregular oscillations.
However, this assumption was not corroborated by observations.

In our preceding paper (Fadeyev 2020) W~Vir pulsating variables were shown to be
the post--AGB stars with masses and luminosities in the ranges
$0.52M_\odot \le M \le 0.55M_\odot$ and
$2\times 10^3L_\odot\lesssim L\lesssim 5\times 10^3L_\odot$,
whereas the fundamental mode period is $\Pi \gtrsim 15$~day.
Bearing in mind observed secular decrease of the period we assume that RU~Cam is in the
early post--AGB stage of its evolution so that observed decrease of oscillation amplitude
is due to the fact that the star evolves beyond the region of pulsation instability.

The goal of the present study consists of two tasks.
First, we determine the mass and the luminosity of the post--AGB stars with
cessation of radial oscillations at the period $\Pi\approx 22$~day.
To this end we use selected models of the evolutionary sequence as initial conditions
for solution of radiation hydrodynamics equations descrinig stellar oscillations and
thereby evaluate the star age, the mass, the luminosity and the radius of the star on the
boundary of pulsation instability.
The pulsation period at the instability boundary is estimated in assumption that the
growth or decay time of oscillations is much shorter than the time of evolutionary changes.
Thus, the system of the equations of hydrodynamics is closed using the steady--state
inner boundary conditions.

Solution of the second task implies construction of the hydrodynamic model describing
decaying radial oscillations during evolution of the star across the boundary of
pulsation instability.
Here the solution of the equations of radiation hydrodynamics is obtained with
time--dependent inner boundary conditions and spans over several dozen years.

\section*{Evolutionary models of RU~Cam}

Stellar evolution from the main sequence up to the early white dwarf stage
was calculated with the program MESA version 12778 (Paxton et al. 2019).
We considered the stars with mass on the main sequence $M_0=0.82M_\odot$, $0.85M_\odot$,
$0.9M_\odot$ and with initial abundances of helium and heavier elements (metals)
$Y=0.25$ and $Z=10^{-3}$, respectively.
Convective mixing was treated according to the theory of B\"ohm--Vitense (1958) with
mixing length to pressure scale height ratio
$\alpha_\mathrm{MLT} = \Lambda/H_\mathrm{P} = 1.8$.
To take into account effects of overshooting we used the presciption by Herwig (2000)
with parameters $f=0.016$ and $f_0=0.004$.
The energy generation rates in thermonuclear reactions and the solution of the equations
of nucleosynthesis were obtained for the reaction network consisting of 26 isotopes
from hydrogen ${}^1\mathrm{H}$ to magnesium ${}^{24}\mathrm{Mg}$ and involving 81 reactions.
The reaction rates were evaluated with the JINA Reaclib data (Cyburt et al. 2010).

Effects of the stellar wind in the stages preceding the AGB phase were treated according to
Reimers (1975) using the parameter $\eta_\mathrm{R} = 0.5$.
Mass loss in the AGB and post--AGB stages of evolution was treated according to Bl\"ocker (1995).
Bearing in mind significant uncertainties in mass loss rates of AGB stars we considered
three parameters in the Bl\"ocker (1995) formula for each value of the initial mass $M_0$:
$\eta_\mathrm{B}=0.02$, 0.05 and 0.1.
The onset of the AGB stage was assumed to correspond to the relative central
helium abundance decreasing below $Y_\mathrm{c} = 10^{-4}$.

Selected models of each evolutionary sequence in the post--AGB stage were used as initial
conditions for calculation of nonlinear stellar pulsations.
Equations of radiation hydrodynamics and time--dependent convection are discussed in our
earlier paper (Fadeyev 2013).
It should be noted that in this section we discuss solution of the hydrodynamics equations
obtained for the steady--state inner boundary conditions
\begin{equation}
\label{ibc}
\frac{\partial r_0}{\partial t} = \frac{\partial L_0}{\partial t} = 0 ,
\end{equation}
where $r_0$ and $L_0$ are the radius and the luminosity at the inner boundary of the
hydrodynamic model.
Using the boundary conditions (\ref{ibc}) we assume that the e-folding time of growing
or decaing oscillations is much smaller than the characteristic time of evolutionary
changes in the structure of the pulsating star envelope.

Main results of stellar evolution and nonlinear stellar pulsation calculations are
summarized in Table~\ref{tabl1}, where for each pair of $M_0$ and $\etab$ characterizing
the evolutionary sequence in the post--AGB stage we give the star age $\tev$,
the stellar mass $M$, the luminosity $L$ and the pulsation period $\Pi$ at the formal
boundary of pulsation instability, that is in the point of the evolutionary track where
the growth rate of kinetic energy of pulsation motions is $\eta=0$ (Fadeyev 2019).

As seen in Table~\ref{tabl1}, three evolutionary sequences
($M_0=0.82M_\odot$, $\eta_\mathrm{B}=0.05$; $M_0=0.82M_\odot$,
$\eta_\mathrm{B}=0.10$; $M_0=0.85M_\odot$, $\eta_\mathrm{B}=0.10$) are consistent with
condition $\Pi\approx 22$~day at the boundary of pulsational instability.
All three models have similar masses and luminosities:
$0.524M_\odot\le M\le 0.532M_\odot$ and $2.20\times 10^3 L_\odot\le L\le 2.330\times 10^3 L_\odot$.
Hydrodynamic models computed for these evolutionary sequences show similar change
of the amplitude of limit--cycle oscillations as the star approaches the boundary
of instability.
The oscillation amplitude of pulsation models of the evolutionary sequence
$M_0=0.82M_\odot$, $\etab=0.05$ is shown in Fig.~\ref{fig1} by dashed lines.
As seen, the significant decrease of the amplitude occurs before the star reaches the
instability boundary.
After its abrupt drop the amplitude of bolometric light variations does not exceed
$\approx 0.1$ mag and slowly decreases to zero for $\approx 18$ yr.

\section*{Hydrodynamic model of RU~Cam}

Hydrodynamic model of RU Cam in the stage of approaching the instability boundary implies
that the steady--state inner boundary conditions (\ref{ibc}) are replaced by explicit
relations $r_0(\tev)$ и $L_0(\tev)$ obtained from the evolutionary computations.
In the present study the inner boundary was set in layers with radiative transfer
in the mass zone with the temperature $T\approx 10^6$\: K and the radius $r_0\approx 0.02R$,
where $R$ is the radius of the upper boundary of the evolutionary model.
The mass of the layers lying above the inner boundary is $\approx 1.1\times 10^{-3}$ of the
stellar mass.

For initial conditions we used the limit--cycle hydrodynamic model of the evolutionary sequence
$M_0=0.82M_\odot$, $\etab=0.05$ locating at $\approx 20$ yr from the formal instability boundary
and pulsating with velocity amplitude of the upper boundary $\Delta U\approx 30$ km/s.
Results of calculations with time--dependent inner boundary conditions are shown in
Fig.~\ref{fig2}, where the maximum kinetic energy $E_{K,\max}$ over the closed pulsation cycle
is plotted as a function of evolutionary time $\tev$.
For the sake of convenience the evolutionary time $\tev$ is set to zero at the formal boundary
of instability determined from calculations with steady--state inner boundary conditions.

As seen in Fig.~\ref{fig2}, the large amplitude oscillations exist for $\tev < -8$ yr and then
for nearly two years the model evolves to small--amplitude oscillations that gradually decay
beyond the formal instability boundary, that is for $\tev > 0$.
Cessation of oscillation decay for $\tev > 15$ yr and the nearly maximum constant
kinetic energy $\log E_{K,\max}\approx 32$ is due to the limited accuracy of
hydrodynamic calculations.

\begin{figure}
\centerline{\includegraphics{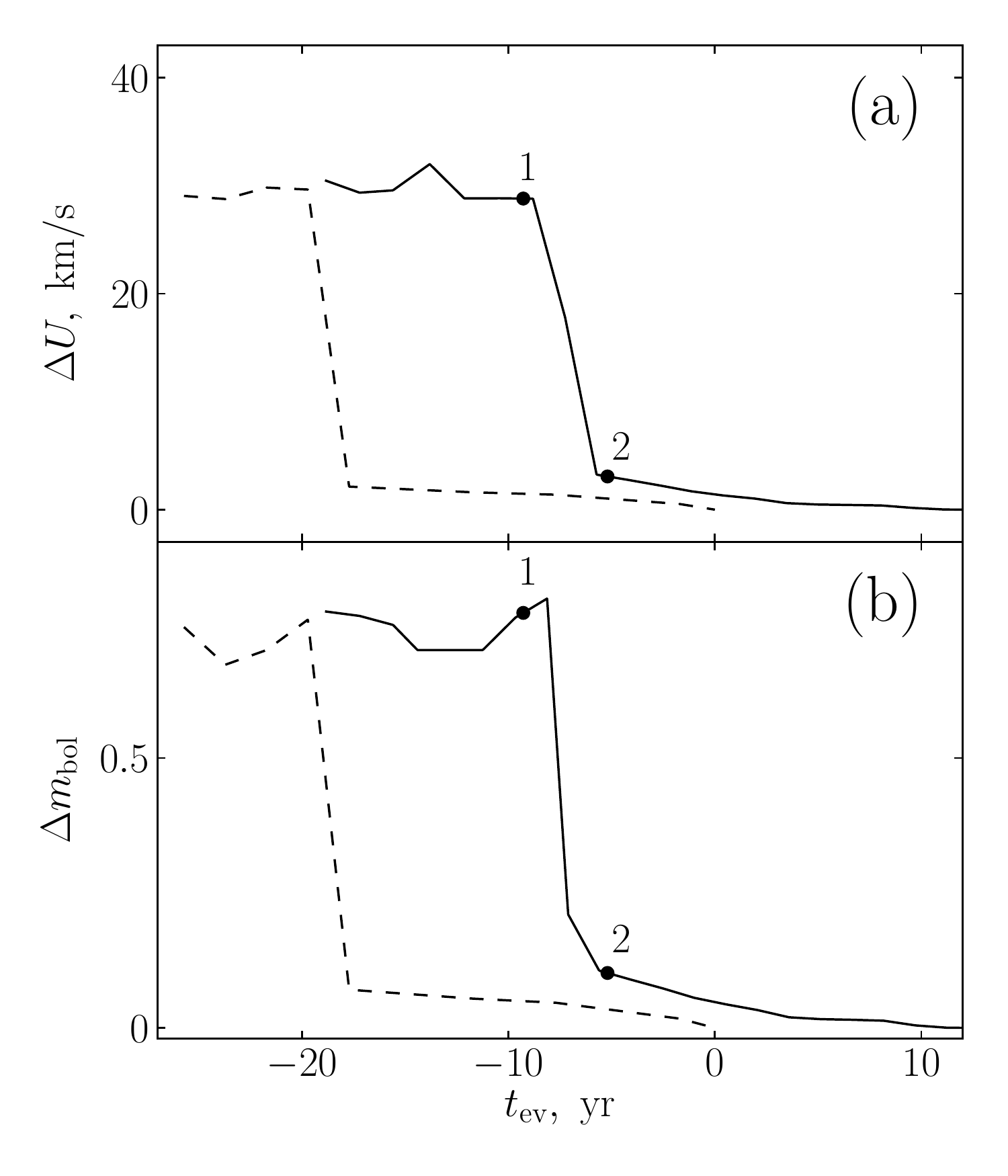}}
\caption{The amplitude of the gas flow velocity at the outer boundary of the hydrodynamic
         model (a) and the amplitude of bolometric light (b) as a function of star age $\tev$
         for the evolutionary sequence $M_0=0.82$, $\etab=0.05$.
         Dashed and solid lines represent the solution obtained with the steady--state and
         the time--dependent inner boundary conditions.}
\label{fig1}
\end{figure}

\begin{figure}
\centerline{\includegraphics{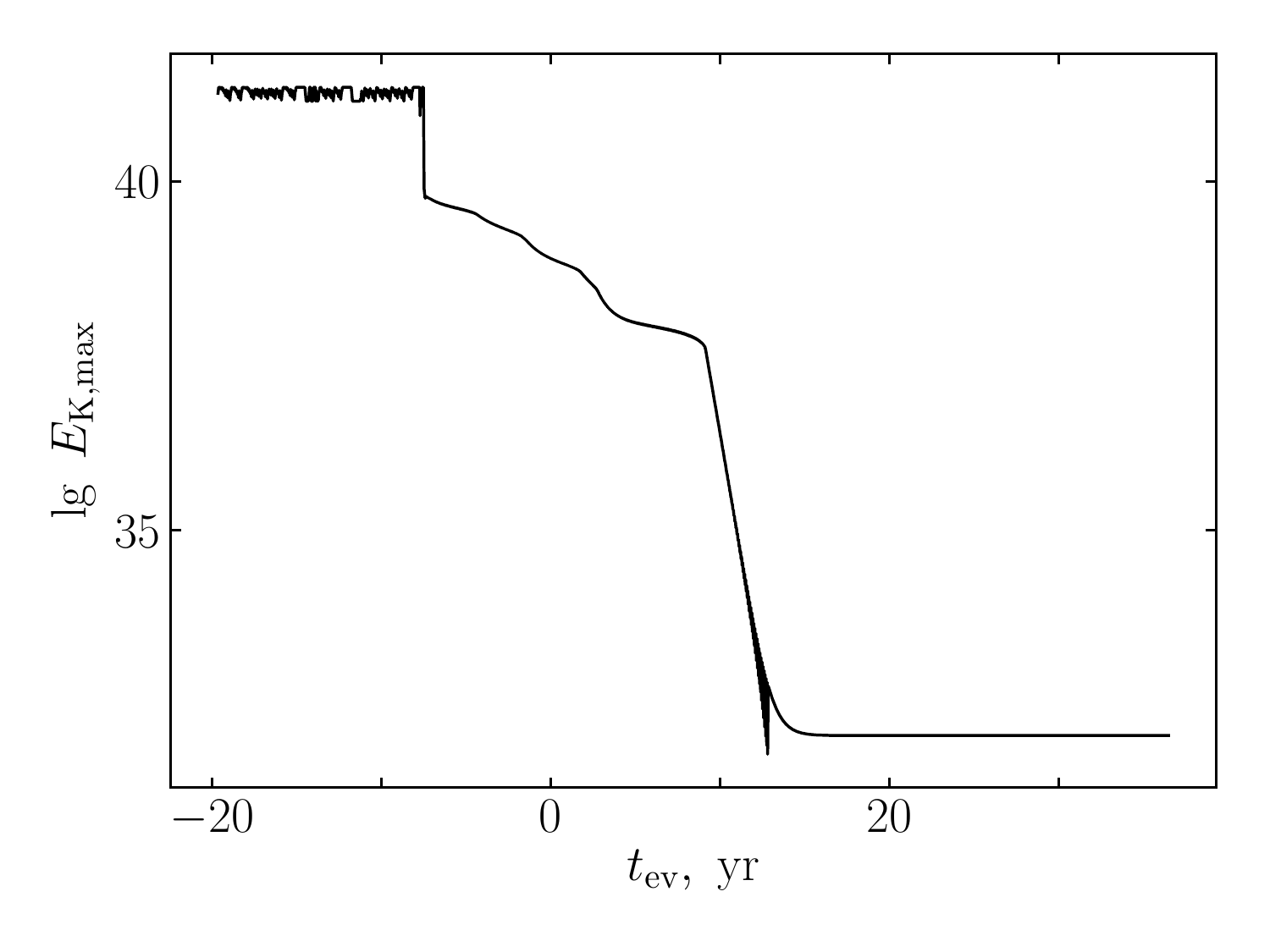}}
\caption{The maximum kinetic energy of the closed pulsation cycle $E_{\mathrm{K},\max}$
         as a function of evolutionary time $\tev$ of the hydrodynamic model with
         time--dependent inner boundary conditions.}
\label{fig2}
\end{figure}

The amplitude of the gas flow velocity variations at the uppermost mass zone $\Delta U$
and the amplitude of the bolometric light $\Delta m_\mathrm{bol}$ computed with
time--dependent inner boundary conditions are shown in Fig.~\ref{fig1} by solid
lines.
Comparison of the plots shown in Fig.~\ref{fig1} by solid and dashed lines
allows us to conclude that the solution obtained with time--dependent boundary conditions
shows significant delay in the large--amplitude limit--cycle oscillations
(oscillation hysteresis) followed by abrupt drop of the pulsation amplitude.

\begin{figure}
\centerline{\includegraphics{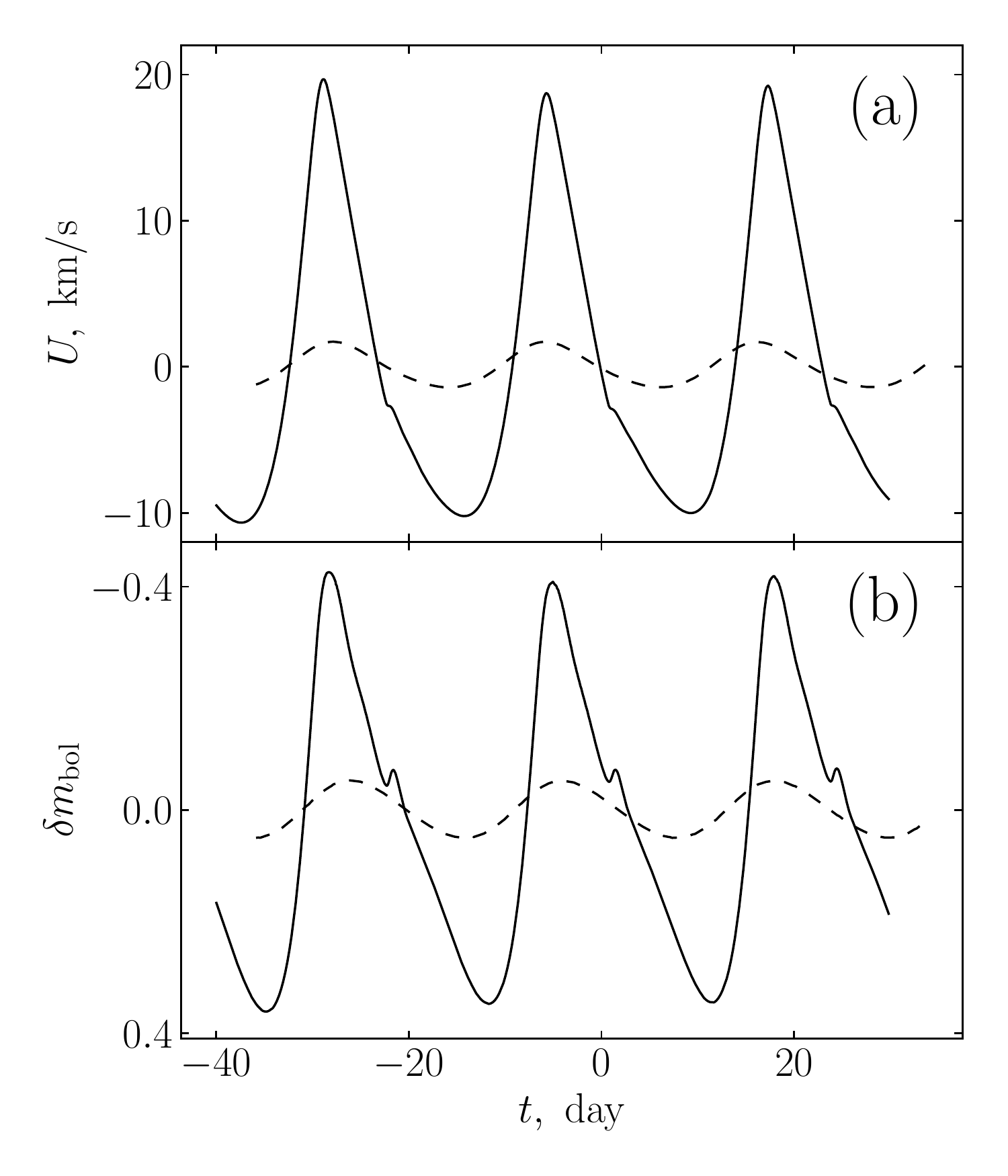}}
\caption{Temporal variations of the gas flow velocity of the uppermost mass zone (a)
         and the bolometric light (b) of the hydrodynamic model.
         Solid and dashed lines correspond to values of the evolutionary time $\tev$
         marked in Fig.~\ref{fig1} by filled circles 1 and 2.}
\label{fig3}
\end{figure}

Variations of the radial velocity at the uppermost mass zone $U$ and the bolometric light
$m_\mathrm{bol}$ are shown for two time intervals before and after the abrupt amplitude drop.
In Fig.~\ref{fig1} they are marked by filled circles as 1 and 2.
As seen from Fig.~\ref{fig3}, nonlinear oscillations of RU~Cam before significant decrease
of the oscillation amplitude are somewhat irregular which is a typical feature of pulsating
post--AGB stars (Fadeyev 2019b).

\section*{Conclusions}

The results of calculations presented above allow us to conclude that abrupt cessation of
large amplitude radial oscillations in RU Cam is due to the fact that the star leaved
the region of pulsational instability and proceeds its evolution across the HR diagram to
higher effective temperatures as the ordinary post--AGB star.
The observed period $\Pi=22$ day allowed us to obtain the theoretical estimates of the
mass and the luminosity for the star on the edge of pulsational instability:
$0.524M_\odot\le M\le 0.532M_\odot$ and $2.20\times 10^3L_\odot\le L\le 2.33\times 10^3L_\odot$.

The use of the time--dependent inner boundary conditions instead of the steady--state
inner boundary conditions has not led to significantly different results of hydrodynamic
calculations.
In both cases the approach of the star to the instability edge is accompanied by significant
($\approx 90\%$) and rapid decrease of the amplitude with following slow decay of
the small--amplitude oscillations.
At the same time, the use of the time--dependent inner boundary conditions showed that
crossing of the pulsation instability boundary is accompanied by the effect of
oscillation hysteresis which leads to displacement of the large--amplitude oscillations
into the region of the small--amplitude oscillations as well as to extension of the stage of the
decaying small--amplitude oscillation beyond the formal instability boundary.

\section*{references}

\begin{enumerate}

\item L.N. Berdnikov and O.V. Voziakova, Inform. Bull. Var. Stars \textbf{4154}, 1 (1995).

\item T. Bl\"ocker, Astron. Astrophys. \textbf{297}, 727 (1995).

\item N. Bogoliubov and Y. Mitropolsky, Asymptotic Methods in the Theory of Non--linear
      Oscillations, 2d ed.; New York: Gordon \& Breach, 1961.

\item E. B\"ohm--Vitense, Zeitschrift f\"ur Astrophys. \textbf{46}, 108 (1958).
      
\item P. Broglia, Inform. Bull. Var. Stars \textbf{213}, 1 (1967).

\item P. Broglia and G. Guerrero, Astron. Astrophys. \textbf{18}, 201 (1972).

\item W. Ceraski, Astron. Nachr. \textbf{174}, 79 (1907).

\item R.H. Cyburt, A.M. Amthor, R. Ferguson, Z. Meisel, K. Smith,
      S. Warren, A. Heger, R.D. Hoffman, T. Rauscher, A. Sakharuk, H. Schatz,
      F.K. Thielemann, and M. Wiescher,
      Astrophys. J. Suppl. Ser. \textbf{189}, 240 (2010).

\item S. Demers and J.D. Fernie, Astrophys. J. \textbf{144}, 440 (1966).

\item Yu.A. Fadeyev, Astron. Lett. \textbf{39}, 306 (2013).

\item Yu.A. Fadeyev, Astron. Lett. \textbf{45}, 353 (2019a).
      
\item Yu.A. Fadeyev, Astron. Lett. \textbf{45}, 521 (2019b).

\item Yu.A. Fadeyev, Astron. Lett. \textbf{46}, 734 (2020).

\item J.D. Fernie, J. Roy. Astron. Soc. Canada \textbf{62}, 214 (1968).

\item F. Herwig, Astron. Astrophys. \textbf{360}, 952 (2000).

\item Z. Koll\'ath and B. Szeidl, Astron. Astrophys. \textbf{277}, 62 (1993).

\item F. Lenouvel and D. Jehoulet, Annales d’Astrophysique \textbf{16}, 139 (1953).

\item A. Michalowska--Smak and J. Smak, Acta Astron. \textbf{15}, 333 (1965).

\item A.V. Nielsen, Astron. Nachr. \textbf{260}, 377 (1936).

\item C. Payne--Gaposchkin, Harvard Bull. \textbf{915}, 10 (1941).

\item B. Paxton, R. Smolec, J. Schwab, A. Gautschy, L. Bildsten, M. Cantiello,
      A. Dotter,  R. Farmer, J.A. Goldberg, A.S. Jermyn, S.M. Kanbur, P. Marchant, A. Thoul,
      R.H.D. Townsend, W.M. Wolf, M. Zhang, and F.X. Timmes,
      Astrophys. J. Suppl. Ser. \textbf{243}, 10 (2019).

\item J.R. Percy, J. Am. Associat. Var. Star Observ. \textbf{49}, 46 (2021).

\item D. Reimers, \textit{Problems in stellar atmospheres and envelopes}
      (Ed. B. Baschek, W.H. Kegel, G. Traving, New York: Springer-Verlag, 1975), p. 229.

\item N.N. Samus', E.V. Kazarovets, O.V. Durlevich, N.N. Kireeva, and E.N. Pastukhova,
      Astron. Rep. \textbf{61}, 80 (2017).
 
\item R.F. Sanford, Astrophys. J. \textbf{68}, 408 (1928).
 
\item B. Szeidl, K. Ol\'ah, L. Szabados, K. Barlai, and L. Patk\'os,
       Commun. Konkoly Observ. \textbf{97}, 245 (1992).

\item G. Wallerstein and D. Crampton, Astrophys. J. \textbf{149}, 225 (1967).

\item C.A. Whitney, Astrophys. J. \textbf{147}, 1191 (1967).

\end{enumerate}

\newpage
\begin{table}
\caption{Parameters of post--AGB stars at the formal boundary of pulsational instability}
\label{tabl1}
\begin{center}
\begin{tabular}{c|c|r|c|c|c}
\hline
 $M_0/M_\odot$ & $\eta_\mathrm{B}$ & $t_\mathrm{ev},10^9$ yr & $M/M_\odot$ & $L/L_\odot$ & $\Pi$, day\\
\hline
 0.82 &  0.02 &  12.450 &  0.537 &  2956 &  30 \\
      &  0.05 &         &  0.528 &  2203 &  22 \\
      &  0.10 &         &  0.524 &  2316 &  23 \\
 0.85 &  0.02 &  10.958 &  0.545 &  2861 &  28 \\
      &  0.05 &         &  0.537 &  2882 &  30 \\
      &  0.10 &         &  0.532 &  2330 &  23 \\
 0.90 &  0.02 &   8.965 &  0.558 &  3309 &  31 \\
      &  0.05 &         &  0.547 &  3071 &  29 \\
      &  0.10 &         &  0.538 &  6667 &  48 \\
\hline
\end{tabular}
\end{center}
\end{table}
\clearpage

\newpage
\clearpage

\newpage
\clearpage

\newpage
\clearpage

\end{document}